\documentclass[conference]{IEEEtran}
\usepackage{cite}
\usepackage{amsmath,amssymb,amsfonts}
\usepackage{graphicx}
\usepackage{textcomp}
\usepackage{xcolor}
\usepackage[hyphens]{url}
\usepackage{fancyhdr}
\usepackage[bookmarks=true,breaklinks=true,letterpaper=true,colorlinks,citecolor=blue,linkcolor=blue,urlcolor=blue]{hyperref}

\usepackage{braket}
\usepackage[inline,shortlabels]{enumitem}
\usepackage[ruled,noend]{algorithm2e}
\usepackage{xfrac}
\usepackage{fdsymbol}

\usepackage{caption}
\usepackage{subcaption}

\usepackage{booktabs}
\usepackage{comment}

\def\BibTeX{{\rm B\kern-.05em{\sc i\kern-.025em b}\kern-.08em
    T\kern-.1667em\lower.7ex\hbox{E}\kern-.125emX}}

\title{Attention-Based Deep Reinforcement Learning for Qubit Allocation in Modular Quantum Architectures}
\author{\IEEEauthorblockN{Enrico Russo}
\IEEEauthorblockA{\textit{University of Catania}\\
Catania, Italy \\
enrico.russo@phd.unict.it}
\and
\IEEEauthorblockN{Maurizio Palesi}
\IEEEauthorblockA{\textit{University of Catania}\\
Catania, Italy \\
maurizio.palesi@unict.it}
\and
\IEEEauthorblockN{Davide Patti}
\IEEEauthorblockA{\textit{University of Catania}\\
Catania, Italy \\
davide.patti@unict.it}
\and
\IEEEauthorblockN{Giuseppe Ascia}
\IEEEauthorblockA{\textit{University of Catania}\\
Catania, Italy \\
giuseppe.ascia@unict.it}
\and
\IEEEauthorblockN{Vincenzo Catania}
\IEEEauthorblockA{\textit{University of Catania}\\
Catania, Italy \\
vincenzo.catania@unict.it}
}
\DeclareMathOperator*{\argmax}{arg\,max}

\SetKwInOut{Input}{Input}
\SetKwInOut{Output}{Output}
\SetKw{Break}{break}
\SetKw{Or}{or}
\SetAlFnt{\footnotesize}

\SetKwComment{Comment}{\#\,}{}
\definecolor{mygreen}{HTML}{004a17}

\SetCommentSty{mycommfont}

\newcommand{\ie}{\textit{i}.\textit{e}., }
\newcommand{\eg}{\textit{e}.\textit{g}., }

\newcommand{\ali}[2]{\makebox[#1][l]{#2}}

\begin{document}

\maketitle
\pagestyle{plain}

%%%%%% -- PAPER CONTENT STARTS-- %%%%%%%%

\begin{abstract}

Modular, distributed and multi-core architectures are currently considered a promising approach for scalability of quantum computing systems. The integration of multiple Quantum Processing Units necessitates classical and quantum-coherent communication, introducing challenges related to noise and quantum decoherence in quantum state transfers between cores. Optimizing communication becomes imperative, and the compilation and mapping of quantum circuits onto physical qubits must minimize state transfers while adhering to architectural constraints. The compilation process, inherently an NP-hard problem, demands extensive search times even with a small number of qubits to be solved to optimality. To address this challenge efficiently, we advocate for the utilization of heuristic mappers that can rapidly generate solutions. In this work, we propose a novel approach employing Deep Reinforcement Learning (DRL) methods to learn these heuristics for a specific multi-core architecture. Our DRL agent incorporates a Transformer encoder and Graph Neural Networks. It encodes quantum circuits using self-attention mechanisms and produce outputs through an attention-based pointer mechanism that directly signifies the probability of matching logical qubits with physical cores. This enables the selection of optimal cores for logical qubits efficiently. Experimental evaluations show that the proposed method can outperform baseline approaches in terms of reducing inter-core communications and minimizing online time-to-solution. This research contributes to the advancement of scalable quantum computing systems by introducing a novel learning-based heuristic approach for efficient quantum circuit compilation and mapping.

\end{abstract}

\section{Introduction}

\begin{figure*}
    \centering
    \includegraphics[width=\textwidth]{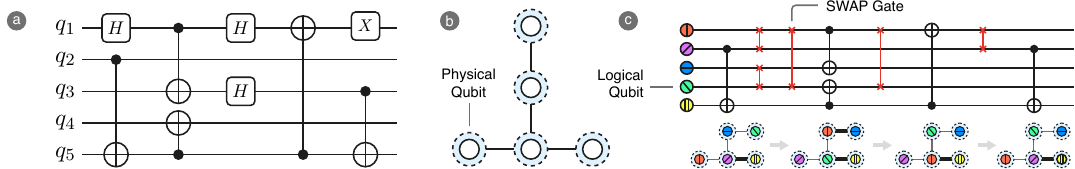}
    \caption{From left to right: a quantum circuit, a quantum processor (IBM Vigo) physical qubit interconnection graph and the resulting circuit with additional SWAP gates after compilation respecting architectural constraint.}
    \label{fig:circuit}
\end{figure*}

Quantum Computing is envisioned as a promising way to solve certain computational tasks exponentially faster than classical computers~\cite{nielsen2010quantum}. Nevertheless, it is widely recognized that for quantum computing to be useful in any real-world problem, a large number of qubits is required~\cite{montanaro2016quantum, shor1999polynomial,grover1996fast}. Hence, the scalability of the quantum system is a strict requirement. Scaling current so-called NISQ (Noisy Intermediate Scale Quantum) systems to large number of qubits is a major challenge due to several issues including, especially for superconducting implementations: technology factors confining the qubits to low fidelity, the need for cryogenic temperatures to reach practical coherence times, the dense integration of control electronic required for each qubit, and crosstalk issues~\cite{charbon2020cryogenic, sarovar2020detecting, arute2019quantum}. The first approach of packing more qubits in a monolithic processing unit within a single large array on the same silicon substrate imposes extremely dense wiring for control electronics, increased crosstalk between qubits and limited yield~\cite{smith2022scaling}. A natural progression inspired by the classical computer architecture evolution is to split the quantum processor into smaller cores~\cite{jnane2022multicore}. The multi-core interconnect should allow classical and quantum communications for quantum state transfer between cores. However, similarly to what happens in the classical counterpart of such architecture, \eg multi-core processors or multi-chiplet systems, inter-core communications are even more expensive then intra-core communications, thus they should be minimized. In the quantum multi-core scenario inter-core quantum state transfer communications are especially expensive in terms of fidelity and coherence-disruptive delays~\cite{rodrigo2021modelling}. 

A quantum circuit comprises a sequence of quantum gate each involving one or more logical qubits. The quantum hardware architecture usually allows only operations involving connected physical qubits. Due to the mentioned technological challenges, physical qubits are sparsely connected. In multi-core systems, the two logical qubits usually needs to be at least allocated in the same core at the moment of the gate execution~\cite{ovide2023mapping}. Hence, for a circuit to be executed on a quantum processor it needs to undergoes transformations that allows to obtain, among the infinite amount of equivalent circuits, one that only includes operations allowed on the target hardware~\cite{nannicini2022optimal}.

This process is known as quantum compilation or quantum transpiling. In this study we target multi-core architecture and in particular we focus on the qubit allocation step. The compilation problem has been demostrated to be NP-complete for single core quantum systems~\cite{botea2018complexity, siraichi2018qubit}. Many methods have been proposed to address the mapping problem in single core architectures. Qubit allocation on multi-core and modular architectures has recently gained attention and few mapping strategies have been proposed~\cite{bandic2023mapping, baker2020time, zhang2023compilation, escofet2023hungarian, escofet2024revisiting}.

We propose a novel technique based on a trending attention-based Deep Reinforcement Learning (DRL) approach to combinatorial problems and sequential decision making, which has been shown to be promosing in many application domains~\cite{kool2018attention,chen2021decision}. We address the peculiarities of the multi-core qubit allocation problem regarding state representation learning leveraging Graph Neural Networks and formulate an appropriate action masking for feasible-only solution construction. We demonstrate the effectiveness of our attention-based agent on mapping random quantum circuits on a 10-core system with grid interconnection topology comparing with black-box optimization baselines. Overall, our study makes the following contributions:
\begin{itemize}
    \item We provide some insights about the design of autoregressive DRL neural agents for the multi-core quantum compilation problem.
    \item We propose a novel attention-based agent able to output feasible solutions with deterministic execution time. This is one of the first attempts for a DRL-based heuristic for multi-core qubit mapping.
    \item We formulate and encode the problem to be solved by several derivative-free optimization algorithms. We then proceed to compare the proposed learned heuristic agent against derivative-free baselines.
\end{itemize}

This paper is organized as follows: Section~\ref{sec:preliminaries} introduces concepts related to quantum circuit and their mapping and provides a formal definition of the problem addressed by this work, Section~\ref{sec:methodology} describes the autoregressive reinforcement learning formulation and the proposed transformer-based policy architecture, and in Section~\ref{sec:experiments} we introduce the priority-based encoding for derivative free optimization algorithms and report the results of the comparison with these algorithms as well as generalization capability assessments.

\section{Preliminaries}
\label{sec:preliminaries}

In this section we formulate the qubit allocation (also known as qubit mapping or qubit partitioning) problem in modular architectures as it will be addressed by the proposed methodology in Section~\ref{sec:methodology}. We start by providing an high-level overview of quantum computing concepts~\cite{nielsen2010quantum, aaronson2013quantum}. We then introduce the qubit allocation problem in modular quantum systems.

\subsection{Quantum Circuits}

In the gate model architecture of quantum computers, quantum programs are expressed as a sequence of reversible gates $G$. Quantum programs are also known as quantum circuits and Fig.~\ref{fig:circuit}a shows an example. Quantum gates specified in a circuit acts on logical qubits and change the quantum state of the qubits.
The state of the qubits is expressed as a linear superposition of basis states as follows:
\begin{equation}
    \ket{\phi} = \alpha\ket{0} + \beta\ket{1} = \alpha \begin{pmatrix} 1 \\ 0 \end{pmatrix} + \beta \begin{pmatrix} 0 \\ 1 \end{pmatrix} = \begin{pmatrix} \alpha \\ \beta \end{pmatrix}
\end{equation}
where $\alpha, \beta \in \mathbb{C}$, hence $\ket{0}$ and $\ket{1}$ are the basis space of a 2D complex vector space and $\ket{\phi} \in \mathbb{C}^2$. A quantum gate can be represented as a unitary matrix that operate on the qubit changing its state. For instance, a common quantum gate is the Hadamaard gate $H$:
\begin{equation}
    \ket{\phi}' = H\ket{\phi} = \frac{1}{\sqrt{2}}\begin{pmatrix} 1 & 1 \\ 1 & -1 \end{pmatrix}\ket{\phi}
\end{equation}
Quantum circuits usually involves more than one qubit and the overall state can be expressed as the tensor product of the state of the qubits\footnote{In absence of entanglement between qubits.}. For instance, for two qubit states $\ket{\phi_1}$ and $\ket{\phi_2}$, the overall state $\ket{\psi}$ is expressed as:
\begin{equation}
\begin{split}
    \ket{\psi} &= \ket{\phi_1} \otimes \ket{\phi_2} = 
    \begin{pmatrix} \alpha_1 & \beta_1 \end{pmatrix}^\top
    \otimes
    \begin{pmatrix} \alpha_2 & \beta_2 \end{pmatrix}^\top
    \\ 
    &=  \alpha_1\alpha_2\ket{00} + \alpha_1\beta_2\ket{01} + \alpha_2\beta_1\ket{10} + \alpha_2\beta_2\ket{11}
\end{split}
\end{equation}
In general $\ket{\psi} \in \mathbb{C}^{2^n}$ where $n$ is the number of qubits in the circuit. Consequently, a quantum gate can act on multiple qubits simultaneously and is represented by a $2^m \times 2^m$ unitary matrix, where $m \leq n$ is the number of qubits it acts on. 

An infinite amount of gates can be represented as unitary matrix operations but only a subset can be practically implemented in quantum hardware. The subset implemented in a quantum computer is considered a \textit{universal set} in the sense that every possible circuit can be reduced to a circuit featuring only gates of this subset. Usually, only 1-qubit and 2-qubit gates are implemented in hardware. The most common 2-qubit gate is named Controlled Not (CNOT). Informally, $\text{CNOT}_{q_1,q_2}$ indicates that $q_2$ is negated when the state of $q_1$ is $\ket{1}$.

In order for a quantum circuit to be executed, after transforming it to a circuit containing only gates available in the hardware, logical qubits have to be mapped on physical qubits, \ie physical objects that behave like a two-state quantum system. The specific mapping location matters especially for the execution of 2-qubit gates. For this to be executed, in fact, a physical connection between the two qubits involved in the operation needs to exist. Thus, at the execution timestep in which, according to the program, the gate is expected to be executed, the logical qubits have to be allocated in two connected physical qubits. In single-core sparsely connected quantum hardware architectures (Fig.~\ref{fig:circuit}b), this is usually achieved by swapping the quantum states (logical qubits) between connected physical qubits as shown in Fig.~\ref{fig:circuit}c. These additional SWAP gates comes with a fidelity cost and introduce noise in the execution.

\subsection{Modular Quantum Systems}

\begin{figure*}
    \centering
    \includegraphics{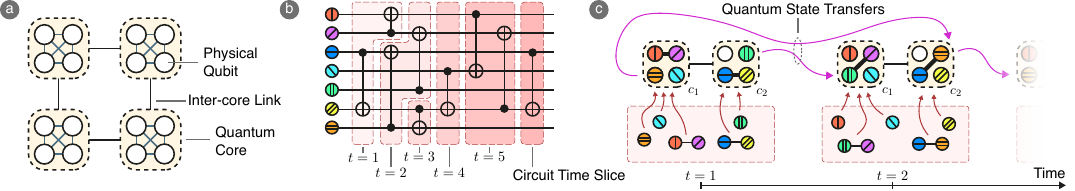}
    \caption{A Multi-core quantum architecture, a sliced quantum circuit and qubit allocation for the first two circuit slices.}
    \label{fig:multicore}
\end{figure*}

Due to technological and physical challenges such as qubit crosstalk, control electronics and cryogenic devices footprint in densely integrated quantum processors and near-perfect yield requirements, the combination of multiple quantum processing units in a modular multi-core system is envisaged as a promising way to scale the number of physical qubits to amounts enabling extensive real-world applications~\cite{smith2022scaling}.

As shown in Fig.~\ref{fig:multicore}a, in a modular quantum system, qubit states are transferred between cores through a quantum link encompassing both classical and quantum communications~\cite{rodrigo2021modelling}. Quantum state teleportation protocols, based on the entanglement phenomenon in quantum systems, can be leveraged in such architectures to transfer a quantum state to another core. Alternatively, remote gates can be applied on qubits allocated on different cores \cite{cuomo2023optimized}. In this work we focus on the first case and we aim to minimize inter-core state transfers.

In this scenario, in order for a quantum gate to be executed, the logical qubits involved in the operation have to be allocated in the same quantum core. In this work, we assume that physical qubits in each core are all-to-all connected, hence the exact physical qubit allocation is not relevant \cite{bandic2023mapping}. Alternatively, we refrain from making any assumptions regarding the interconnectivity of the quantum cores (Fig.~\ref{fig:multicore}a).

\subsection{Circuit Slicing and Qubit Allocation}

A common first step in qubit allocation optimization both in single-core \cite{nannicini2022optimal} and multi-core \cite{baker2020time,bandic2023mapping,escofet2024revisiting} architectures is to split the circuit to be mapped in \textit{slices} that contains only gates that can be executed in parallel, \ie gates not sharing any of the logical qubits they act on, as shown in Fig.~\ref{fig:multicore}b. In modular architectures, the next step, for each circuit slice, consists in allocating logical qubits in quantum cores making sure that: \begin{enumerate*}[(a)] 
\item \textit{friends} qubit, namely qubit involved in the same gate, are allocated in the same core;
\item the amount of logical qubits allocated in each core does not exceed its capacity, namely the number of physical qubits\footnote{We refer to physical \textit{computation} qubits. In multi-core architecture some of the physical qubits of the cores might be reserved for quantum state transfer protocols~\cite{rodrigo2021modelling,zhang2023compilation}.} the core has.
\end{enumerate*}
In performing this process, we want to minimize the number of inter-core communications needed to move the logical qubits from the core in which they are allocated in a time slice, to the cores where they are allocated in the next slice. In sparsely connected multi-core architectures, state transfer between cores can have different costs depending on the hop distance.

\subsection{Problem Formulation}
\label{sec:formulation}
Formally, the problem just described addressed in this study is a combinatorial optimization problem that can be formulated as:
\begin{align}
\min_{x}\quad & \omit\rlap{$\displaystyle\sum_{t=1}^{T-1}\sum_{q=1}^Q\sum_{s=1}^C\sum_{d=1}^C x_{t,q,s}x_{t+1,q,d}D_{s,d}$} \label{eq:obj} \\
\text{s.t.}\quad 
        & x \in \{0,1\} &&  \nonumber \\
        & \textstyle \sum_{c=1}^C x_{t,q,c} = 1  &&\forall q \quad \forall t  \label{eq:onecore}\\
        &\textstyle\sum_{q=1}^Q x_{t,q,c} \leq P_c &&\forall t    \label{eq:capacity}\\
&\textstyle\sum_{c=1}^C x_{t,g_1,c}x_{t,g_2,c} \geq 1 && \forall (g_1,g_2) \in G_t \quad \forall t \label{eq:friendship}
\end{align}
where $x$ are the binary decision variables, $x_{t,q,c} = 1$ if at timestep $t$ the $q$-th logical qubit is allocated in the $c$-th core, $T$ is the number of time slices of the circuit, $Q$ is the number of logical qubits, $C$ is the number of cores, $D_{s,d}$ is the state transfer cost or distance between $s$-th and $d$-th cores, $P_c$ is the capacity of the $c$-th core, $G_t$ is the set of logical qubits pairs representing 2-qubit gates\footnote{Similar formulation can be obtained for gates acting on more qubits. In this work we focus on 2-qubit gates as they are more frequently implemented in quantum architectures \cite{barenco1995elementary}.} in the $t$-th slice, Eq.~\eqref{eq:onecore} constraints each logical qubit to be allocated in exactly one core in each time slice, Eq.~\eqref{eq:capacity} takes into account core capacities, Eq.~\eqref{eq:friendship} enforces in each slice and for each gate that qubits involved are allocated in the same core, and the objective in Eq.~\eqref{eq:obj} is the amount of inter-core communications.

To obtain the partitioning of a quantum circuit in time slices the iterative Algorithm~\ref{alg:slicing} can be considered. The algorithm operates on a given list $G$ of 2-qubit gates, represented as pairs of logical qubit indices. For each gate $(q_1, q_2)$, the process entails starting from the last time slice, moving backward until a slice $t$ is encountered that contains gates involving either qubit $q_1$ or $q_2$. Subsequently, the gate is appended to slice $t+1$. If no such slice is identified, the gate is added to the first slice. If slice $t+1$ does not exist, it is created and appended to the output list of slices denoted as $S$.

\begin{algorithm}
\caption{Quantum circuit slicing procedure}\label{alg:slicing}
\Input{\ $G = [(g^1_1,g^1_2),\ldots,(g^L_1,g^L_2)]$ \Comment*[r]{list of gates}} 
\Output{\ $S=[G_1,\ldots,G_T]$ \Comment*[r]{list of slices}}
$T \gets 0$\;
$S \gets []$\;
\ForEach{$(q_1,q_2) \in G$}{
\For{$t \leftarrow T$ \KwTo $-1$}{
    \If{$t = -1$ \Or $q_1 \in \bigcup S[t]$ \Or $q_2 \in \bigcup S[t]$}{
        \If{$t+1 > T$}{
            $T \gets t + 1$\;
            $S$.append($\varnothing$)
        }
        $S[t+1] \gets S[t+1] \cup \{(q_1,q_2)\}$\;
        
        \Break
        }
    }
}

\end{algorithm}

\subsection{RL for Combinatorial Optimization}

RL serves as a fundamental framework within the field of artificial intelligence, offering a paradigm that enables machines to learn and make decisions through interaction with an environment. At its essence, RL involves an agent, responsible for taking actions within a given context (state), and an environment, which responds by providing rewards or penalties based on these actions. The primary goal of the agent is to acquire an optimal policy, a strategic mapping of states to actions that maximizes cumulative rewards over time. This learning process, inspired by human-like trial-and-error mechanisms, has proven remarkably successful in diverse applications including robotics, control, games, autonomous driving and many others~\cite{li2023survey,wang2020deep,sutton2018reinforcement,giannelli2022tutorial}. 

Recent works have demonstrated promising results of RL applied to combinatorial optimization problems \cite{bello2016neural,berto2023rl4co}. As the solution space of combinatorial grows exponentially, it quickly become infeasible for a neural network to output a feasible solution in one shot, due to the gargantuan action space. In our case, the number of possible decision variable assignments (including non-feasible solutions) for the problem described in Eq.~\eqref{eq:obj} is equal to $2^{TQC}$, \eg about $10^{9030}$ when mapping a 30-slice circuit of 100 qubits on a 10-core architecture. To address these issues, similarly to language models \cite{vaswani2017attention}, autoregressive methods that generates feasible solution step-by-step taking into account constraints have been recently proposed \cite{berto2023rl4co,kool2018attention}. In such methods the problem $\mathbf{x}$ is first encoded using a trainable encoder $f_\theta$, obtaining an encoded representation $\mathbf{h}$:
\begin{equation} \label{eq:encoder}
    \mathbf{h} = f_\theta(\mathbf{x})
\end{equation}
then, the trained policy decoder $g_\theta$ outputs the best action probability distribution at each timestep $t$, based on the encoded representation of the problem and the current partial solution (result of the past actions), formally we express this as follows:
\begin{equation}
    a_t \sim g_\theta(a_t|a_{t-1},\ldots,a_0,\mathbf{h})
\end{equation}
at each timestep we select an action that keeps the solution valid, eventually obtaining a complete solution. Thus, given the problem $\mathbf{x}$, the policy $\pi_\theta$ outputs a probability distribution for a solution $\mathbf{a}$ built in $T$ decoding steps:
\begin{equation}
    \pi_\theta(\mathbf{a}|\mathbf{x}) \triangleq \prod_{t=1}^T g_\theta(a_t|a_{t-1},\ldots,a_0,f_\theta(\mathbf{x}))
\end{equation}
According to the RL paradigm training the policy means finding the set of optimal parameters $\theta^*$ that maximize the expected return, namely the expected value of the reward given the distribution of policy actions and of the problem instances, formally \cite{berto2023rl4co}:
\begin{equation}
    \theta^* = \argmax_\theta\  \mathbb{E}_{\mathbf{x}\sim P(\mathbf{x})} \!\left[ \mathbb{E}_{\mathbf{a}\sim \pi_\theta (\mathbf{a}|\mathbf{x}) } \!\left[ R(\mathbf{a},\mathbf{x})\right] \right]
\end{equation}
In the following section we describe how we applied this paradigm to the qubit allocation task in multi-core quantum architectures.

\section{Methodology}
\label{sec:methodology}

In this section we describe the proposed methodology for RL-based heuristic for the multi-core qubit allocation problem. Initially, we furnish an outline of encoding and decoding procedures in the trained policy. Subsequently, we introduce the components essential to represent and encode the input circuit and the environment state. Finally, we go deeper describing each component and we describe how we make sure that the output solution is a valid solution.

\subsection{Autoregressive RL for Qubit Allocation}

Autoregressive RL has been successfully applied to many routing problems such as the Travelling Salesman Problem (TSP) and the Capacited Vehicle Routing Problem (CVRP) in \cite{kool2018attention}, which also introduced the attention-based model policy. For instance, in TSP the input is a set of coordinates and output is a permutation of these that signifies the node traversal order. In such problems, a valid solution is obtained leveraging Pointer Networks \cite{vinyals2015pointer} concepts and at each decoding step the next city (action) is selected among the input nodes, masking already visited nodes. Thus, such problems only require one encoder for the input nodes.

Recently, such methods have been applied to other domains each featuring specific constraints and challenges, such as machine scheduling \cite{chen2022deep}, user-server allocation \cite{chang2023attention} and hardware design \cite{kim2023devformer}. Applying the autoregressive RL paradigm to qubit allocation problem also comes with unique challenges.

%The first important feature is that the possible actions are not represented by the input sequence as in the TSP problem \cite{kool2018attention}, for this reason we adopt a dual-encoder approach. Another peculiarity is that, differently from \cite{chang2023attention}, 

\begin{algorithm}
\caption{Policy workflow}\label{alg:policy}
\Input{\ $X=[G_1,\ldots,G_T]$ \Comment*[r]{seq. of slices}}
\Output{\ $A=[A_1,\ldots,A_T]$ \Comment*[r]{alloc. $\forall$ slice}}
$A \gets []$\;
$\mathbf{H}^{(I)} \gets$ \texttt{InitEmbedding($X$)}\;
$\mathbf{H}^{(S)} \gets$ \texttt{EncoderBlocks($\mathbf{H}^{(I)}$)}\;
$\mathbf{H}^{(X)} \gets$ \texttt{AveragePooling($\mathbf{H}^{(S)}$)}\;
%$\mathbf{h}^Q \gets$ \texttt{GenerateEmbeddings()}\;
\For(\Comment*[f]{for each circuit slice}){$t \leftarrow 1$ \KwTo $T$}{
    $A_t \gets []$\;
    $\mathbf{H}^{(C)}_t \gets$ \texttt{SnapshotEncoder($A_{t-1}$)}\;
    \For(\Comment*[f]{for each logical qubit}){$q \leftarrow 1$ \KwTo $Q$ }{ 
        context $\gets$ \texttt{concat($\mathbf{H}^{(X)}$,$\mathbf{H}^{(S)}_t$,$\mathbf{E}^{(Q)}_q$)}\;
        {$\mathbf{f}$} $\gets$ current free capacities\;
        {$\mathbf{d}$} $\gets$ distance from $q$'s core in $t-1$\;
        {$\mathbf{G}^{(C)}_{t,q}$} $\gets$  \texttt{DynamicEmbedding($\mathbf{H}^{(C)}$, $\mathbf{f}$, $\mathbf{d}$)}\;
        \ali{1em}{$a$} $\gets$ \texttt{MaskedPointer(}context, $\mathbf{G}^{(C)}$\texttt{)}\;
        $A_t$.append($a$)\;
    }
    $A$.append($A_t$)\;
}

\end{algorithm}

The proposed overall workflow of the trained policy receiving an input quantum circuit and returning as output a valid solution is summarized in Algorithm~\ref{alg:policy}. The policy receives a sequence of circuit slices as input. Each slice $t$ is appropriately encoded by the \texttt{InitEmbedding} component to obtain representation embeddings $\mathbf{H}^{(I)}_t\in\mathbb{R}^{d_E}$, where $d_E$ is the embedding size. Similarly to \cite{kool2018attention}, several transformer \texttt{EncoderBlock}s are employed to achieve a slice representation that takes into account relations with all the other slices in the input circuit.

Subsequently, the decoding process starts. The policy outputs the core choice for each logical qubit for each slice, sequentially. Thus, the number of decoding steps is equal to $T \cdot Q$, \ie number of slices in the circuit times the number of (used) logical qubits. The decoding process can be considered 3-level hierarchical. In fact, the context based on which the policy takes the decision for the next action is three-fold and consists in: circuit representation $\mathbf{H}^{(X)}$, slice representation $\mathbf{H}^{(S)}_t$, logical qubit representation $\mathbf{E}^{(Q)}_q$. The circuit representation remains the same across the decoding process for a particular input, the slice representation remains the same for $Q$ decoding steps, while the qubit representation changes at each decoding timestep.

The action space at each decoding timestep, \ie for each logical qubit allocation, is represented by the possible cores in which the qubit can be allocated. Differently from the routing problems employing a single encoder model, we employ another encoder that we name \texttt{SnapshotEncoder} to obtain a representation of the qubit allocation in the previous circuit slice. In particular, an embedding $\mathbf{H}^{(C)}_c$ is obtained for each core $c$ that incorporates information about the logical qubits allocated in it and the relationship with other cores. These embeddings are calculated when moving from the decoding of one slice to the next. Furthermore, for each timestep, namely for each logical qubit, additional dynamic information is incorporated in core embeddings. In particular for each core $c$, information about the current available capacity and information about the cost to transfer the current logical qubit to $c$ from the core in which it was allocated in the previous slice.

Given the three-fold context and the augmented core embeddings, at each decoding step a masked attention-based pointer mechanism is employed to select the core for logical qubit. The mask makes sure that the qubit can be allocated only in cores that keep the solution valid and allows to complete a valid solution in the next timesteps.

After all the decoding timesteps, for each slice we obtain the core on which each logical qubit is allocated. Furthermore, the obtained solutions will respect the problem constraints.

\subsection{Circuit Slice Encoder}

\begin{figure}
    \centering
    \includegraphics{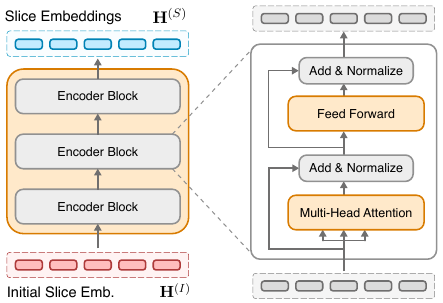}
    \caption{Attention-based circuit slice encoder.}
    \label{fig:encoder}
\end{figure}

The first step of the autoregressive policy is obtaining a learned representation of the input circuit as in Eq.~\eqref{eq:encoder}. Traditional formulations of RL problems as Markov Decision Process (MDP), expect the agent to takes decision considering a representation of the current state which is only the result of the past state and the past actions. In the autoregressive paradigms applied to combinatorial optimization, we have the chance to first build a global representation of the input and only then sequentially consider contexts that focus on a part of it, as in the MDP. In our case, we design an encoder that allows to build a representation of each slice incorporating global information (circuit-level information). Fig.~\ref{fig:encoder} depicts the circuit slice encoder of the proposed policy model. We achieve an embedding for each circuit slice employing the attention-based encoder blocks proposed in \cite{vaswani2017attention}.

\subsubsection{\texttt{InitEmbedding}}

\begin{figure}
    \centering
    \includegraphics{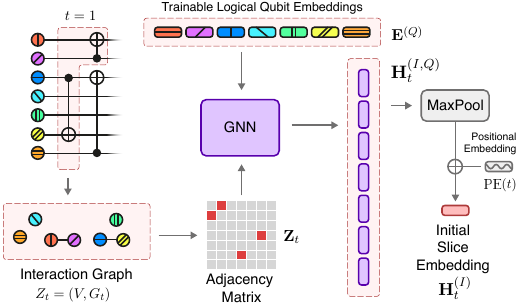}
    \caption{Initial embedding of the time slice $t=1$ through a GNN layer.}
    \label{fig:initembed}
\end{figure}

  In this regard, the challenge specific to the qubit allocation problem is that we first need to represent the input of the transformer encoder as a set of initial embeddings $\mathbf{H}^{(I)}_t\in\mathbb{R}^{d_E}$ for each slice $t$ where $d_E$ is the size of such embedding. The output of the slicing Algorithm~\ref{alg:slicing}, described previously, is the sequence of the $T$ slices we need to embed. Each slice $G_t$ consist in a set of logical qubit pairs representing the gates contained in the slice. We can consider $G_t$ as the edge set of a undirected disconnected graph $Z_t = (V,G_t)$ representing the interaction between logical qubits $V$ given by gates \cite{bandic2023mapping}. Because of this, we decide to embed each slice employing a Graph Neural Network (GNN), as shown in Fig.~\ref{fig:initembed}. In particular we employ the convolutional graph operator introduced in \cite{kipf2016semi}. In our case, for each slice $t$, we obtain a qubit-wise representation $\mathbf{H}^{(I,Q)}_{t} \in \mathbb{R}^{Q \times d_E}$ as follows:
\begin{equation}
    \mathbf{H}^{(I,Q)}_{t} = \mathbf{\tilde{D}}^{-\frac{1}{2}}_t \mathbf{\tilde{Z}}_t
\mathbf{\tilde{D}}^{-\frac{1}{2}}_t \mathbf{E}^{(Q)} \mathbf{W}^{(I)},
\end{equation}
where $\mathbf{W}^{(I)}$ is a trainable weight matrix, \mbox{$\mathbf{E}^{(Q)} \in \mathbb{R}^{Q \times d_E}$} are trainable logical qubit embeddings, $\mathbf{\tilde{Z}}_t \in \{0,1\}^{Q \times Q}$ is the adjacency matrix of the interaction graph $Z_t$ with added self-loops and $\mathbf{\tilde{D}}^{-\frac{1}{2}}_t$ the element-wise reciprocal square root of its diagonal degree matrix. On the obtained qubit-wise embeddings $\mathbf{H}^{(I,Q)}_{t}$ we apply a max pooling across the qubit dimension $Q$. Then, sinusoidal positional encoding \cite{vaswani2017attention} is applied finally obtaining the embedding \mbox{$\mathbf{H}^{(I)}_{t} \in \mathbb{R}^{d_E}$} for each circuit time slice $t$.

\subsubsection{\texttt{EncoderBlocks}}
The initial slice embeddings $\mathbf{H}^{(I)} \in \mathbb{R}^{T\times d_E}$ are then processed by $b$ transformer encoder block \cite{vaswani2017attention,kool2018attention,chang2023attention}, that allows to incorporate in each slice embedding, information from other relevant circuit slices. Ideally, this is beneficial because in selecting core for qubits of a slice considering its embedding, the policy can be aware of the next circuit slices. Furthermore, the transformer model surpasses the limitations associated with conventional neural sequence transduction models, including recurrent and convolutional neural networks. Its enhanced capability to grasp long-range dependencies within a sequence, positions it as a superior solution for addressing large-scale problems. Moreover, the transformer model's efficacy is particularly advantageous in the context of parallel execution. Each encoder block consist in a Multi-Head Attention layer (MHA) followed by a fully connected (FC) layer. Both are followed by a residual connection and a batch normalization layer. We now introduce the attention mechanism and the multi-head attention variant because it will be of interest in the decoding phase as well. For a complete description of the encoder block, the interested reader can refer to \cite{vaswani2017attention}.
The attention function consist in the calculation of an output as sum of input value vectors weighted by the \textit{compatibility} of the input query vector with input key vectors corresponding to each value. The scaled-dot-product attention function is defined as follows:
\begin{equation}
    \text{Attention}(\mathbf{Q}, \mathbf{K}, \mathbf{V}) = \text{softmax}\left(\frac{\mathbf{QK}^\top}{\sqrt{d_K}}\right)\mathbf{V}
\end{equation}
where $\mathbf{Q}$ are the query vectors, $\mathbf{K}$ are the key vectors and $\mathbf{V}$ value vectors. $d_K$ is the dimensionality of the key vectors. $\mathbf{Q}$, $\mathbf{K}$ and $\mathbf{V}$ are obtained through 3 different fully connected layers (linear projections) applied on some input. If the queries $\mathbf{Q}$ and the key-value pairs $\mathbf{K}$, $\mathbf{V}$ are calculated starting (projecting) from the same input vectors, the attention is named self-attention. If $\mathbf{Q}$ is calculated on a different input, it is called cross-attention.
The MHA calculation consists in combining the result of the attention from different attention \textit{heads} as follows:
\begin{align}
\label{eq:attention}
\textbf{H}_i &= \text{Attention}(\mathbf{Q}\mathbf{W}^Q_i, \mathbf{K}\mathbf{W}^K_i, \mathbf{V}\mathbf{W}^V_i) \\
    \text{MHA}(\mathbf{Q}, \mathbf{K}, \mathbf{V}) &= \text{concat}(\textbf{H}_1,\ldots,\textbf{H}_h)\textbf{W}^O
\end{align}
where $\mathbf{W}$ are linear projection trainable weights and $h$ is the number of heads. In the first encoder the MHA is applied in self-attention mode to the projections of the initial slice embeddings, incorporating in each slice embedding information from other \textit{compatible} slices. Subsequent encoder blocks perform the same operation on the output of the previous encoder block. The output of the $b$-th (last) encoder block represents the ultimate slice embeddings $\mathbf{H}^{(S)}\in\mathbb{R}^{T\times d_H}$, where $T$ is the number of circuit slices and $d_H$ is the dimensionality of the embeddings, considered in the decoding phase.

\subsection{Core Snapshot Encoder}

\begin{figure}
    \centering
    \includegraphics{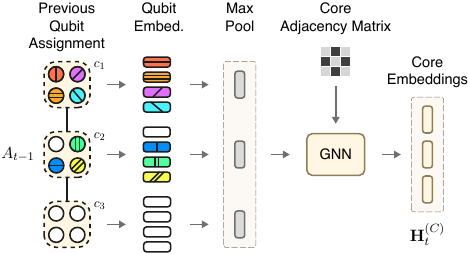}
    \caption{Previous slice qubit assignment snapshot encoder.}
    \label{fig:snapshot}
\end{figure}

The task of the core \texttt{SnapshotEncoder} is to provide an encoding of the qubit allocations in the previous circuit time slice. During the decoding process, in fact, we also want the policy to take into account the previous qubit allocation, possibly trying to make a compromise between the state transfer distances from the previous slices and the qubit allocation for the next slices still to be decided.

During the decoding, when processing the $t$-th time slice, the output action for the previous slice $t-1$ is a vector \mbox{$A_{t-1} \in \{1,\ldots,C\}^Q$}, \ie it consists in the mapping between each logical qubit and the core index. As shown in Fig.~\ref{fig:snapshot}, the Core Snapshot Encoder transform the allocation it represents in the embeddings $\mathbf{H}_t^{(C)}\in\mathbb{R}^{C\times d_H}$. This is obtained employing another GNN similarly to the \texttt{InitEmbedding} component. This time the input adjacency matrix is fixed and represents the connectivity of the multi-core hardware architecture. Regarding the node input features instead, these are obtained for each core by performing the max pooling of the embeddings of the logical qubits allocated on it. Logical qubit embeddings are calculated for each physical qubit available in the core. If a physical qubit is not associated to logical qubit, a padding embedding is considered.

\subsection{Decoding Process}

\begin{figure}
    \centering
    \includegraphics{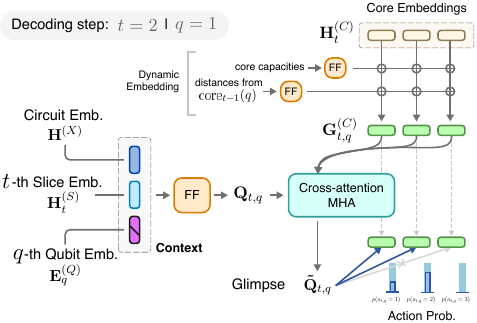}
    \caption{Action probability calculation at each decoding step.}
    \label{fig:decoding}
\end{figure}

As already described in previous sections and summarized in Algorithm~\ref{alg:policy}, the decoding process is carried for $T\times Q$ decoding steps. Each logical qubit of each slice is allocated sequentially. Thus, a decoding step can be identified by the indices pair $(t,q)$ indicating the current slice and the current logical qubit being considered respectively. At each decoding step, the context is represented by the concatenation of three embeddings: \begin{enumerate*}[(a)] 
\item the circuit embedding $\mathbf{H}^{(X)}$ calculated as the average of the slice embeddings obtained from the circuit slice encoder;
\item the slice embedding $\mathbf{H}^{(S)}_t$ currently being considered;
\item the trainable embedding $\mathbf{E}^{(Q)}_q$ of the logical qubit currently being allocated.
\end{enumerate*}

When starting the decoding process for a new time slice $t$, the qubit allocation snapshot of the previous slice $t-1$ is encoded in the embeddings $\mathbf{H}^{(C)}_t\in\mathbb{R}^{C\times d_H}$ by the core \texttt{SnapshotEncoder}. As shown in Fig.~\ref{fig:decoding}, for each logical qubit $q$ being allocated, these emeddings are augmented with useful information and this procedure is called \texttt{DynamicEmbedding}. In particular, given the current remaining capacity for each core, these are individually projected to embeddings with the same dimensionality $d_H$ and then summed to $\mathbf{H}^{(C)}_t$. The current remaining capacity of each core is reset to its number of physicaal qubits when the allocation of a new slice $t$ begins, namely every $Q$ decoding steps. When a qubit is allocated to a core, its capacity is decreased by 1. Furthermore, a similar process is carried out to incorporate the distance of the qubit $q$ being allocated, from each core, based on the core where it was allocated in the previous slice and the architecture core distance matrix. These distances are projected at each timestep to an embedding with the same dimensionality $d_H$ and summed to the core embeddings. Finally, from this \texttt{DynamicEmbedding} procedure, we obtain core embeddings \mbox{$\mathbf{G}^{(C)}_{t,q}\in\mathbb{R}^{C\times d_H}$} that incorporate information about: \begin{enumerate*}[(a)] 
\item qubit allocation in the previous slice;
\item current remaining capacity;
\item hypothetical state transfer cost for qubit being mapped;
\end{enumerate*}

As shown in Fig.~\ref{fig:decoding}, the final step of each decoding step is obtaining the probabilities of allocating the qubit $q$ of slice $t$ on each possible core $c$, starting from the current context and the core embeddings $\mathbf{G}^{(C)}_{t,q}$. This task is achieved with an attention mechanism \cite{kool2018attention} and it first involves projecting the context to obtain a query vector as follows:
\begin{equation}
    \mathbf{Q}_{t,q} = \text{concat}\left(\mathbf{H}^{(X)},\mathbf{H}^{(S)}_t,\mathbf{E}^{(Q)}_q\right)\mathbf{W}^{(V,G)}
\end{equation}
where $\mathbf{W}^{(V,G)}$ is the trainable weight matrix for query projection from the context. Then, the cross attention is performed between the query and the core embeddings obtaining the attended query vector $\mathbf{\tilde{Q}}_{t,q}$ also known as \textit{glimpse} \cite{bello2016neural,vinyals2015order}:
\begin{align}
     \mathbf{K}_{t,q} &= \mathbf{G}^{(C)}_{t,q}\mathbf{W}^{(K,G)} \\
      \mathbf{V}_{t,q} &= \mathbf{G}^{(C)}_{t,q}\mathbf{W}^{(V,G)} \\
    \mathbf{\tilde{Q}}_{t,q} &= \text{MHA}\left(\mathbf{Q}_{t,q},\mathbf{K}_{t,q}, \mathbf{V}_{t,q}\right)
\end{align}
where $\mathbf{W}^{(K,G)}$ and $\mathbf{W}^{(V,G)}$ are the trainable weights matrix projecting key and value vectors from the augmented core embeddings. Subsequently, compatibilities are calculated between the attended query and the core embeddings, similarly to the attention function in Eq.~\eqref{eq:attention}. Compatibilities with cores on which mapping the current logical qubit $q$ would result in an invalid solution are masked, thus having:
\begin{equation}
\label{eq:logit}
    u_{t,q,c} =
\begin{cases}
    -\infty,              & \text{if masked}\\
    \frac{\mathbf{\tilde{Q}}_{t,q}\mathbf{K}_{t,q}^\top}{\sqrt{d_K}},& \text{otherwise}
\end{cases}
\end{equation}
The resulting vector $\mathbf{U}_{t,q}$ can be interpreted as logarithm of probabilities (logits) and the final output action probability for the $(t,q)$  decoding timestep can be computed using the softmax function:
\begin{equation}
    p(a_{t,q}=c) = \frac{e^{u_{t,q,c}}}{\sum_{j=1}^C e^{u_{t,q,j}}} \quad\quad \text{for}\ c=1,2,\dots,C
\end{equation}
Next section describes when a core (action) is masked and cannot be selected.

\begin{figure*}
     \centering
     \begin{subfigure}[b]{0.19\textwidth}
         \centering
         \includegraphics[height=130pt]{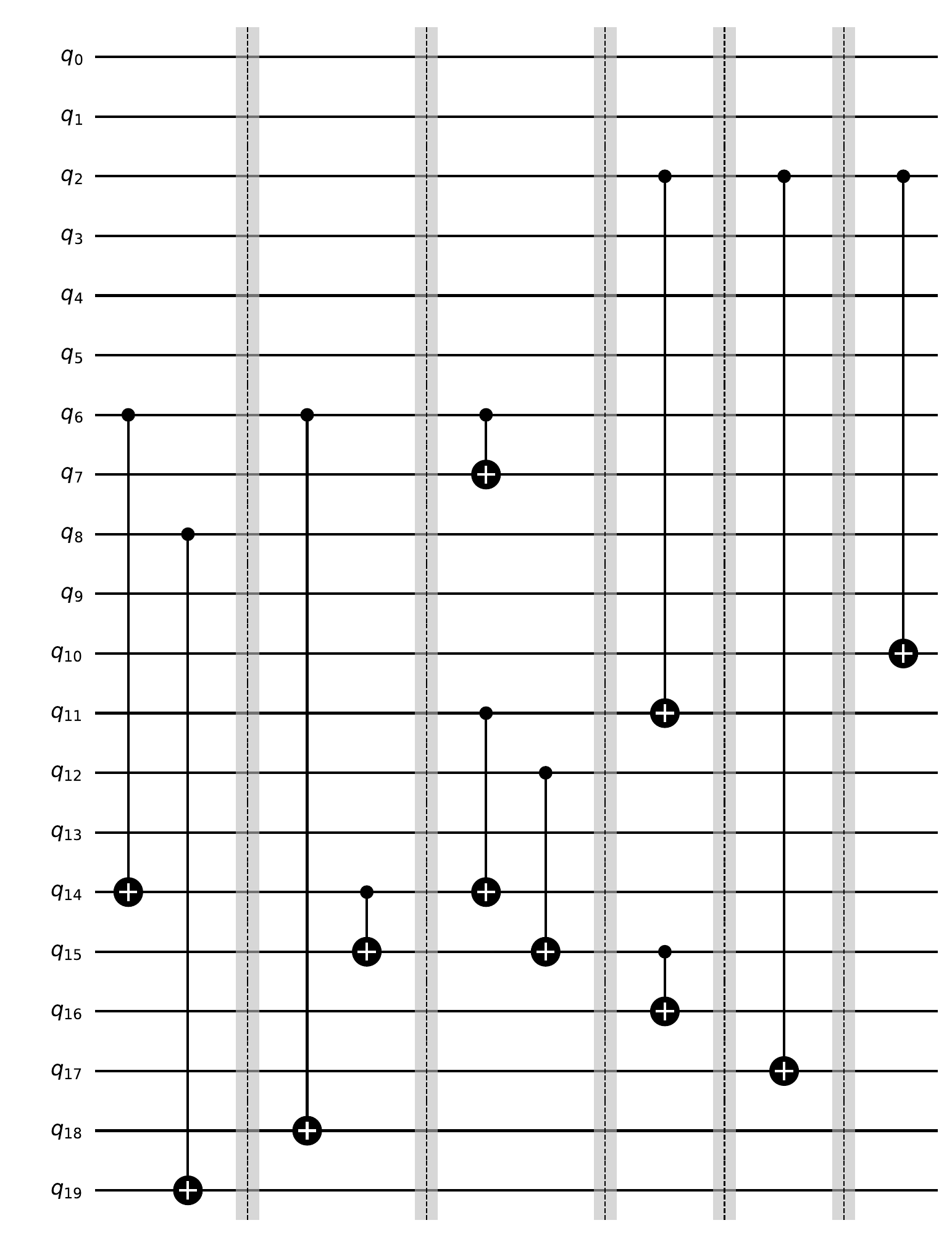}
         \caption{Sliced quantum circuit.}
         \label{fig:mapped_circuit}
     \end{subfigure}
     \hfill
     \begin{subfigure}[b]{0.38\textwidth}
         \centering
         \includegraphics[width=\textwidth]{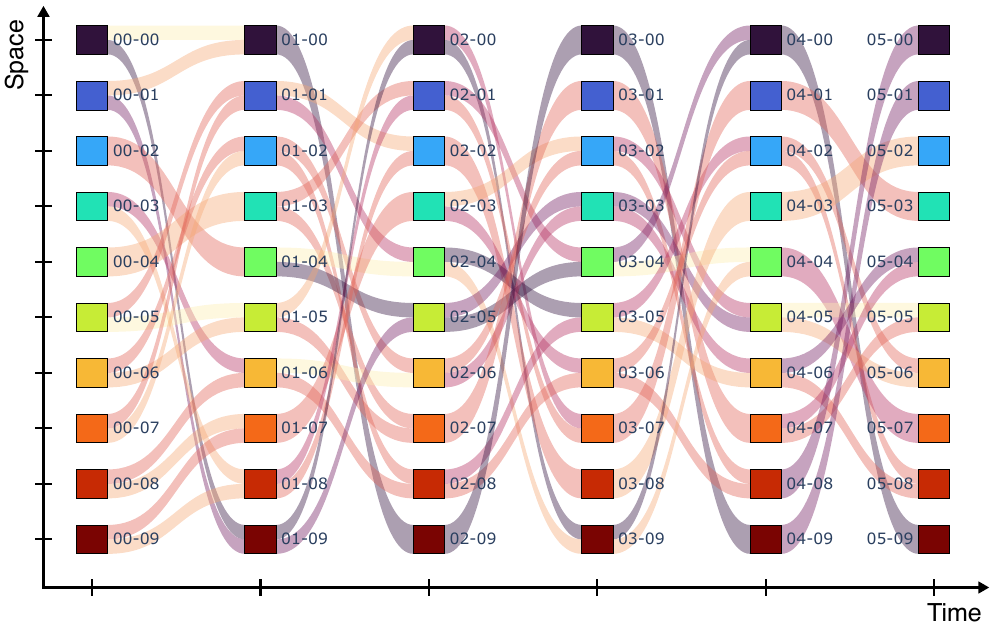}
         \caption{Untrained policy qubit allocation.}
         \label{fig:untrained_policy}
     \end{subfigure}
     \hfill
     \begin{subfigure}[b]{0.38\textwidth}
         \centering
         \includegraphics[width=\textwidth]{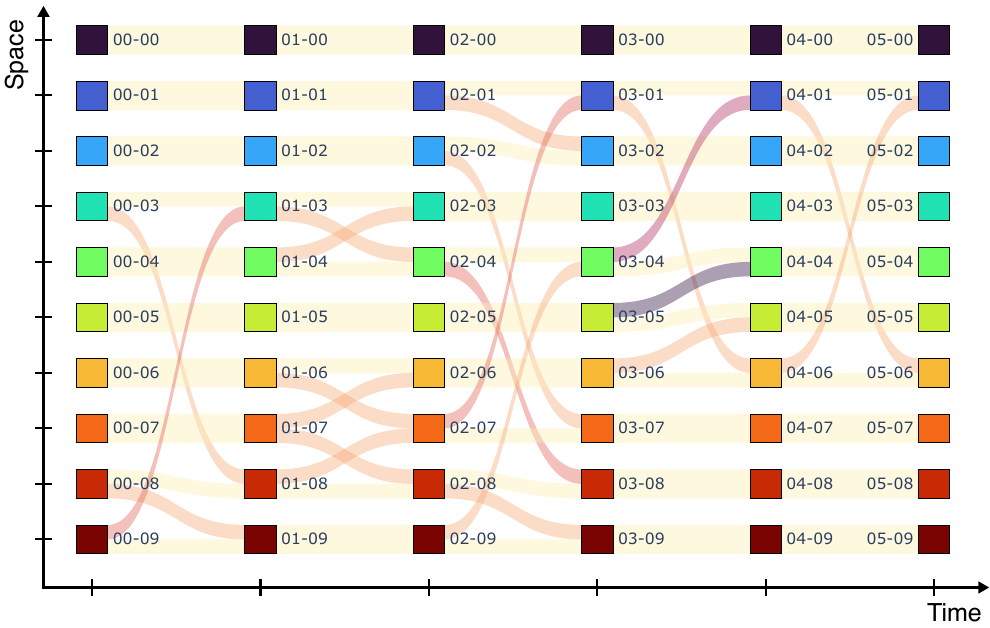}
         \caption{Trained policy qubit allocation.}
         \label{fig:trained_policy}
     \end{subfigure}
        \caption{Inter-core state transfers visualization of two allocation strategies for a 20-qubit 6-slice quantum circuit on a \mbox{10-core $\times$ 2-qubit} system featuring a grid topology. In the two Sankey diagrams, each row represent a core, each column is a time slice. A link represents a quantum state transfer. Darker links signify state transfer between distant cores in the grid.}
        \label{fig:three graphs}
\end{figure*}

\subsection{Action Masking}

Appropriately masking logits in Eq.~\eqref{eq:logit} is essential to build a valid solution. At each decoding step, action masking is based on the current partial solution, result of the previous action choices. 

The first and most straightforward masking regards the constraint in Eq.~\eqref{eq:capacity}, namely the core capacity constraint. A core logit should be masked if its current remaining capacity is $0$ and hence no other qubits can be allocated on it in the current time slice. 

Another necessary masking mechanism is relevant for the constraint in Eq.~\eqref{eq:friendship} relevant to logical qubits \textit{friendship} determined by gate interactions. During the decoding process logical qubits are allocated sequentially from $q=1$ to $q=Q$, when two qubits $q_1<q_2$ are involved in the same gate, $q_1$ is allocated first. The core in which $q_1$ is allocated is taken into account when the action for $q_2$ is being decoded and all the cores different from the $q_1$'s core are masked. An issue can arise if, allocating gates $q_1<q_i<q_2$, the capacity of the $q_1$'s core is exhausted before $q_2$ is allocated. For this reason, when allocating a qubit $q_1$ involved in a gate, the capacity of the core is reduced by $2$ reserving the spot for the interacting qubit $q_2$. When $q_2$ is allocated, no remaining capacity is reduced.

There is still one issue to be addressed that we describe using an example. Given an architecture featuring 2 cores $(c_1,c_2)$ each having 2 physical qubits, and a circuit with four qubits $q_1$,$q_2$,$q_3$, and $q_4$, consider a slice in which \mbox{$q_3\leftrightblackspoon q_4$} are involved in the same gate while $q_1$ and $q_2$ are non-interacting. In the decoding process, $q_1$ and $q_2$ are allocated before \mbox{$q_3\leftrightblackspoon q_4$} and it may happen that \mbox{$q_1 \rightarrow c_1$} and \mbox{$q_2 \rightarrow c_2$}. At this point each of the two cores has a remaining capacity of $1$, and completing a feasible solution is not possible because \mbox{$q_3\leftrightblackspoon q_4$} cannot be allocated in the same core. To avoid this scenario, when allocating a qubit, the remaining amount of interacting couples to be mapped $g$ is taken into account, and only cores in which allocating the qubit would result in \mbox{$\sum_{c=1}^C\lfloor\text{capacity}_c / 2\rfloor\geq g$} are not masked.

\subsection{Reward and Training}

According to the problem formulation in Sec.~\ref{sec:formulation}, after building an overall solution sampling the probability distributions at each decoding step, the corresponding reward is calculated as the total number of inter-core communications needed to bring a qubit in its assigned core from the core where it was allocated in the previous circuit slice. Formally, the reward is defined as:
\begin{equation}
    \label{eq:reward}
    R(A) = \sum_{t=2}^{T}\sum_{q=1}^Q D_{A_{t-1,q},A_{t,q}}
\end{equation}
where $D$ is the distance matrix and $A_{t,q}$ is the core (action) selected for $q$-th qubit at $t$-th slice. The training loss is then defined as the expected value of $R(A)$ and optimized by gradient descent using the REINFORCE \cite{williams1992simple} gradient estimator with rollout baseline as in \cite{kool2018attention}.

\section{Evalutation}
\label{sec:experiments}

In this section we provide implementation details and evaluation results of the proposed methodology.

\subsection{Experimental Setup}
\label{sec:setup}

We implemented the proposed method using PyTorch~\cite{pytorch}, PyTorch~Geometric~\cite{pytorch_geometric} and RL4CO~\cite{berto2023rl4co} libraries. Our models are trained considering two architectures with $10$ cores each featuring $10$ physical qubits. One architecture features an all-to-all core connectivity (\textit{A2A}) and the other features $2\times5$ mesh grid topology (\textit{Grid}). We assume that physical qubits in each core are all-to-all connected as the inter-core communication is more costly then intra-core communication. Furthermore, we train our models on randomly generated quantum circuits of $30$ time slices and $50$ or $100$ logical qubits. Hence, we obtain and test $4$ trained models considering the following combinations: \textit{A2A-50}, \textit{A2A-100}, \textit{Grid-50}, \textit{Grid-100}.
All of them were trained for $100$ epochs of $10,240$ randomly generated samples (circuits) each. Batch size was set to $128$ for $100$ qubits and $256$ for $50$ qubits due to memory constraints. The embedding dimensionality $d_E$ and the latent space size $d_H$ were set to $256$ in all the policy components. The number of attention heads in all MHAs is $8$. The amount $b$ of encoder blocks in the policy is $3$. Learning rate was set to $10^{-4}$ and Adam optimizer was employed. Training was performed on a NVIDIA RTX 3060 12GB GPU leveraging CUDA Automatic Mixed Precision. On our system training process employed about $5$ and $10$ hours to complete for $50$ and $100$ qubit policy versions respectively.

Fig.~\ref{fig:untrained_policy} shows a visualization of inter-core state transfers as a result of the qubit allocation for the circuit in Fig.~\ref{fig:mapped_circuit} performed by the proposed policy untrained model with randomly initialized weights, namely the result of a valid random qubit allocation. In Fig.~\ref{fig:trained_policy} we qualitatively show the improvement for the same circuit over a random qubit allocation due to the training process the policy undergoes.

\subsection{Iterative black-box optimization approaches}

\begin{figure*}
     \centering
     \begin{subfigure}[b]{0.24\textwidth}
         \centering
         \includegraphics[width=0.95\textwidth]{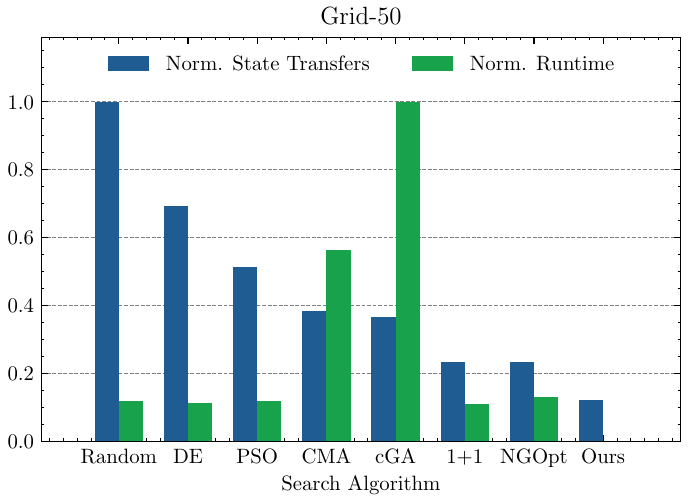}
     \end{subfigure}
     \hfill
     \begin{subfigure}[b]{0.24\textwidth}
         \centering
         \includegraphics[width=0.95\textwidth]{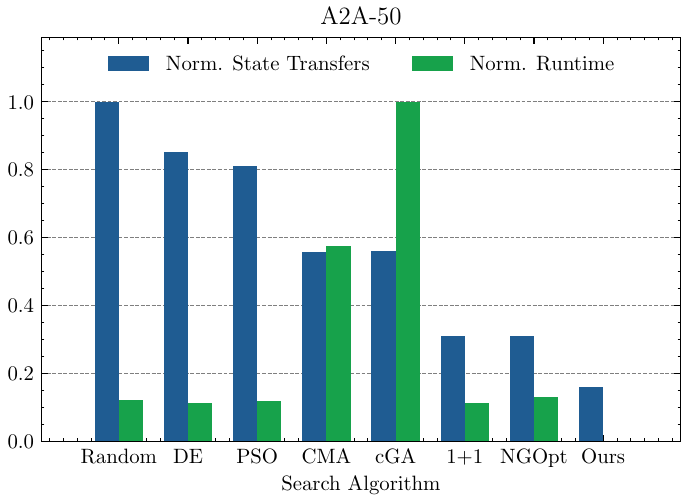}
     \end{subfigure}
     \hfill
     \begin{subfigure}[b]{0.24\textwidth}
         \centering
         \includegraphics[width=0.95\textwidth]{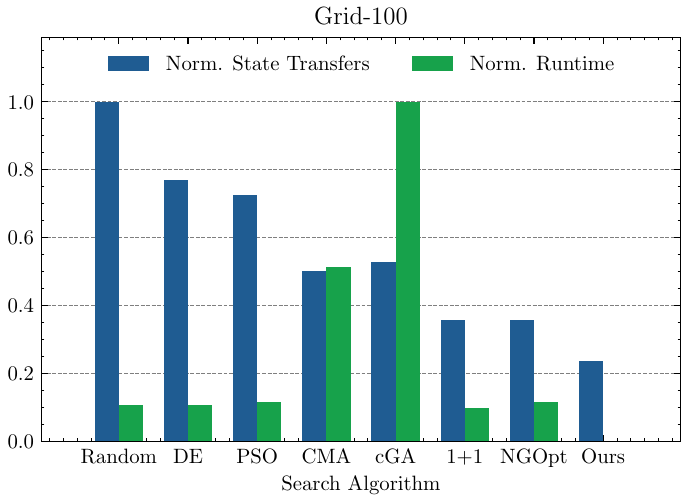}
     \end{subfigure}
     \hfill
     \begin{subfigure}[b]{0.24\textwidth}
         \centering
         \includegraphics[width=0.95\textwidth]{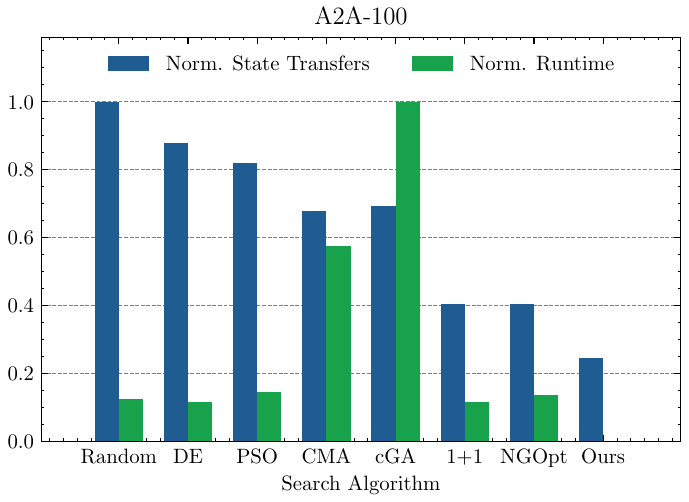}
     \end{subfigure}
        \caption{Inter-core communication count and runtime comparison between the trained policy and black-box iterative optimization approaches on \textit{Grid} and \textit{A2A} architectures.}
        \label{fig:blackbox}
\end{figure*}

In this section, we report the results of the comparison between mapping solution obtained using iterative derivative-free optimization approaches and the proposed trained policies. In order to solve the qubit allocation problem using off-the-shelf iterative algorithms such as genetic algorithms, we need an appropriate solution encoding on which the algorithm operators (such as crossover and mutation operators for genetic algorithm) can work to create new possibly better solutions in terms of a fitness function. We avoid a solution encoding that would comprise infeasible solutions as this would jeopardize the sampling efficiency of the considered baselines. We rather adopt a common approach in literature, also used to solve the Travelling Salesman Problem using genetic algorithms \cite{gen1997genetic,nowling2011priority}, consisting in a priority-based encoding scheme.

In our case the solution is encoded with a real-valued array \mbox{$\mathbf{x} \in \mathbb{R}^{T \times P}$} where $T$ and $P$ are the number of slices of the circuit and the total number of physical qubits in the multi-core system. 
In the decoding of each slice, we intially assume that the amount of logical qubits is equal to the number of physical qubits. Starting from a priority array $\mathbf{x}_t \in \mathbb{R}^{P}$ and considering each core to have a capacity $C$, the first $C$ qubits with the highest priority are allocated to the first core, the second $C$ qubits to the second and so on. To avoid generating non valid solutions, the position of the qubit with the highest priority of a couple involved in a gate determines also the core in which its \textit{friend} is allocated. Finally, only the first $Q \leq P$ qubits, used in the circuit, are considered in the output solution.

The result solution fitness can be evaluated according to the same reward function considered for the proposed methodology in Eq.~\ref{eq:reward}. This encoding scheme allows us to test several gradient-free optimization algorithms: Random search, Differential evolution with two points crossover (DE), Particle Swarm Optimization (PSO), Covariance matrix adaptation evolution strategy (CMA), Compact Genetic Algorithm (cGA) \cite{harik1999compact}, One Plus One evolutionary algorithm (1+1) \cite{droste2002analysis} and NGOpt, a Nevergrad-specific adaptive optimization algorithm. The results of the comparison with the proposed methodology are shown in Fig.~\ref{fig:blackbox}. We considered two random 30-slice circuits with 50 and 100 logical qubits and assessed the result of the mapping on both the \textit{Grid} and \textit{A2A} topology with the respective trained policy. The sampling budget for the iterative optimization algorithms was set to $10^6$ solutions and default parameters provided by the library were used. Iterative algorithm search time ranged from 30 minutes to more than 4 hours for each circuit, while the proposed trained policy returned an heuristic solution in seconds. Furthermore, across the four scenarios the proposed methodology achieved inter-core communications savings ranging from $33.5\%$ to $48.5\%$ with respect to the best performing baseline approach.

\subsection{Generalization capabilities}
\begin{figure}
    \centering
    \includegraphics[width=0.8\linewidth]{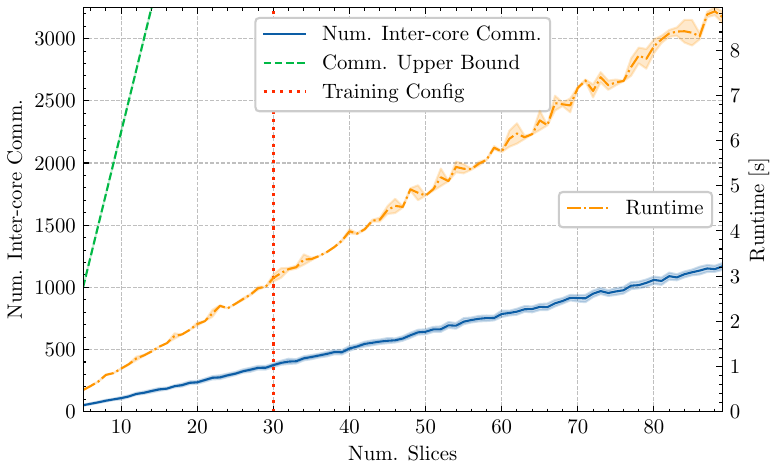}
    \caption{Number of inter-core communications and allocation runtime when mapping random 50-qubits circuits with a variable number of slices using the \textit{Grid-50} policy.}
    \label{fig:generalization}
\end{figure}

\begin{figure}
    \centering
    \hspace*{-0.17in}
    \includegraphics[width=0.75\linewidth]{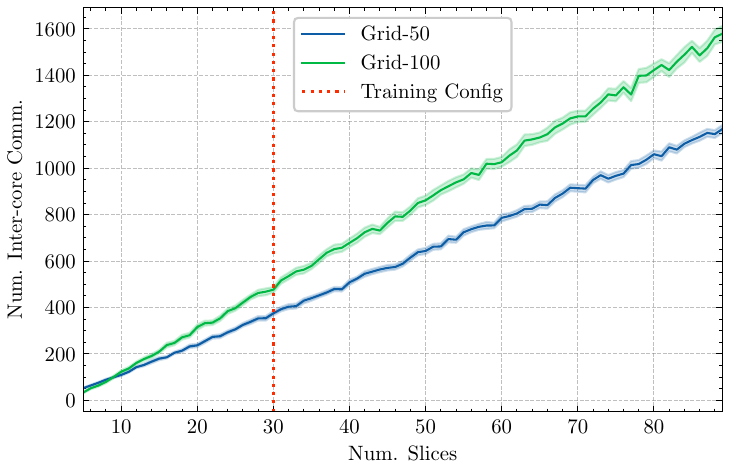}
    \caption{Comparison of inter-core communications when mapping random 50-qubits circuits using the \textit{Grid-50} and \textit{Grid-100} policies trained on 50-qubit and 100-qubit 
circuits respectively.}
    \label{fig:generalization_qubit}
\end{figure}

In Fig.~\ref{fig:generalization} we report the performance of the proposed policy trained on circuits with $50$ qubits and $30$ slices on circuits with a variable number of slices different from the amount on which the policy was trained on. The dashed curve represents the upper bound in terms of number of state transfers for different numbers of circuit slices ranging from $5$ to $90$. The upper bound $\text{ub}(T,Q)$ is calculated assuming that each qubit is transferred to the farthest core at each slice, hence it is calculated as: \[\text{ub}(T,Q) = (T-1)\times Q \times \max{\mathbf{D}}\]
where $T$ is the number of slices, $Q$ the number of qubits and $\mathbf{D}$ the distance matrix. For this experiment we considered the 10-core grid topology.

We notice that there is no evident degradation in the number of inter-core communications as it grows linearly with the number of slices. Hence, we can assume good generalization capabilities of the model to circuit of variable depths. It still remains to be assessed the quality with respect to the optimal solution for different circuit sizes. This task is challenging due to the amount of decision variable involved, which cause current available solvers to run for days to found the optimal qubit allocation even for a small amount of circuit slices \cite{gurobi}.

In Fig.~\ref{fig:generalization}, the runtime in seconds is also reported for our setup (Sec.~\ref{sec:setup}). It can be noticed that for the circuit depths considered, the runtime grows linearly up to 
about $8$ seconds for $90$ slices. The linear growth is due to the slice-wise decoding process. Notice that the reported runtimes refer to a single inference, but it should be taken into account that in the same amount of time a consumer GPU can execute multiple circuit allocations in parallel. For our setup due to available GPU memory, we were able to run $64$ allocations in parallel up to circuits with $60$ slices and $32$ allocations in parallel up to $90$ slices. In any case, these values should be considered device specific. Furthermore, scaling to even higher numbers of slices would be possible dividing the circuit in smaller chunks with a tractable number of slices and mapping each individual chunk. Depending on the size of the circuit, communication overhead generated at junctions between chunks might be negligible. The worst case overhead can be calculated as $\text{ub}(K,Q)$ where $K$ is the number of chunks in which we split the circuit.

We trained the models on random circuit with a fixed amount of used logical qubits, namely 50 and 100. However, the trained models can be used to map circuits with a number of used qubits less or equal the number on which they were trained. We now assess the generalization capability about the number of used qubits by comparing the 50 and 100 qubit version of the policy on 50-qubit random circuits. The results are shown in Fig.~\ref{fig:generalization_qubit}. For 30-slice circuits, same slice count using during training for both, the \textit{Grid-100} result in mappings with about $24\%$ more state transfers, but the overhead grows bigger as the number of slices increases. Training the policies with circuits with variable number of used qubits, might be a way to increase generalization capability of the model.

\begin{comment}
    
\subsection{Optimality for small circuits}

We were able to find the optimal qubit allocation solution for the problem formulated in Eq.~\ref{eq:obj} only for very small problem instances using Gurobi solver \cite{gurobi}. In particular we considered two 50 qubit random circuits with 4 and 8 time slices, mapped on the grid architecture. The former was solved in few seconds giving a zero-communication solution, against the heuristic solution given by the proposed policy featuring $X$ state transfers. The latter took more than 12 hours to be solved resulting in 16 communications against the X of the \textit{Grid-50} trained policy.

\end{comment}

\subsection{Comparison with state-of-the-art}

In this section we analyze the performance of the trained policies on benchmark quantum circuits. We evaluate the performance for the \textit{Grid} and \textit{A2A} case. For the latter we compare the results with a state-of-the-art technique for multi-core qubit allocation named Fine Grained Partitioning Overall Extreme Exchange (FGP-OEE) and the relaxed version (FGP-rOEE) \cite{baker2020time}. 
\begin{table}
  \caption{Trained policy performance for the $2\times5$ multi-core topology on benchmark circuits.}
  \label{tab:circuits}
  \begin{center}
  \footnotesize
  \begin{tabular}{l c c c | c}
    \toprule
     \textbf{Benchmark} & \begin{tabular}{@{}c@{}}\textbf{Num.} \\ \textbf{Qubits}\end{tabular} & \begin{tabular}{@{}c@{}}\textbf{Num.} \\ \textbf{Slices}\end{tabular} & \begin{tabular}{@{}c@{}}\textbf{Dual Gate} \\ \textbf{Count}\end{tabular} & \begin{tabular}{@{}c@{}}\textbf{Inter-core} \\ \textbf{Comm. Count}\end{tabular} \\
    \midrule
    QFT & 50 & 99 & 1225  & 1361 \\
     & 100 & 197 & 4950  & 8918 \\
    \midrule
    Quantum & 50 & 50 & 1226  & 1892 \\
    Volume & 100 & 100 & 4961  & 11438 \\
    \midrule
    Graph State & 50 & 83 & 596  & 827 \\
     & 100 & 166 & 2449  & 5382 \\
    \midrule
    Draper Adder & 50 & 120 & 925  & 1049 \\
     & 100 & 245 & 3725  & 7406 \\
    \midrule
    Cuccaro Adder & 50 & 290 & 336  & 250 \\
     & 100 & 590 & 686  & 1782 \\
    \midrule
    QNN & 50 & 195 & 2498  & 1627 \\
     & 100 & 395 & 9998  & 10438 \\
    \midrule
    Deutsch-Jozsa & 50 & 49 & 49  & 125 \\
     & 100 & 99 & 99  & 570 \\
    \bottomrule
  \end{tabular}
  \end{center}
\end{table}

These two techniques support fully connected (\textit{A2A}) core topologies only. For the proposed technique, in Table~\ref{tab:circuits} the amounts of inter-core communications for benchmark circuits when compiled for the \textit{Grid} architecture (sparse topology) using the trained policies are reported. QFT is the Quantum Fourier Transform without final qubit reordering swaps. Graph State refers to the encoder circuit for a graph having as many nodes as qubits in the circuit (namely $50$ or $100$ in the table) and a random adjacency matrix with $0.5$ density. Drapper and Cuccaro Adder are fixed precision adders of two quantum state registers with half the amount of logical qubits of the circuit. QNN represent a Quantum Neural Network parametrized circuit equivalent to classical Neural Networks. Deutsch-Jozsa algorithm is one of the first proposed algorithm with exponential speed up with respect to classical algorithm, based on a black box function which takes as input a binary string and outputs a bit, it determines whether the function is constant or balanced (the output is 0 for half of the inputs and 1 for the other half). All the circuits were first optimized and decomposed in two-qubit gates using Qiskit framework~\cite{Qiskit}. For the QNN and Deutsch-Jozsa algorithms, the MQT Benchmark Library was considered~\cite{quetschlich2023mqtbench}. For the other circuits, we adopted the Qiskit circuit library implementations.

\begin{figure}
    \centering
    \includegraphics[width=0.8\linewidth]{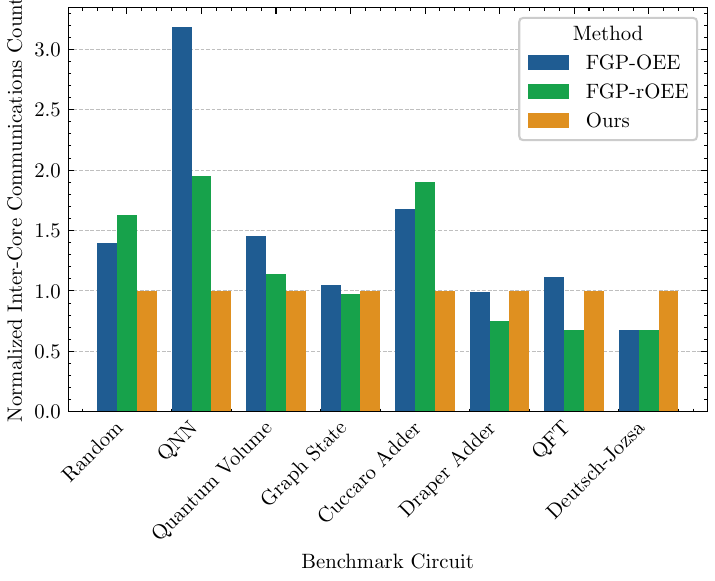}
    \caption{Number of inter-core communication in \textit{A2A-100} architecture for different benchmark circuits normalized with respect to the output of the proposed technique.}
    \label{fig:fgp}
\end{figure}

Fig.~\ref{fig:fgp} shows the results of the comparison of the proposed technique with respect to FGP-OEE \cite{baker2020time}. The same circuits in Table~\ref{tab:circuits} were considered, but compiled using the trained policy for the \textit{A2A} multi-core architecture. The first step of the FGP-OEE method consists in building a \textit{lookahead} interaction graph for each time slice with weighted edges. At timestep $t$, the weight of the edge connecting two qubits $q_i$ and $q_j$ is calculated taking into account how near in the future (next timeslices) the vertexes interact, using the following equation:
\begin{equation*}
    w_t(q_i,q_j) = \sum_{t < m \leq T} I(m,q_i,q_j)\cdot2^{t-m}
\end{equation*}
where $I(m,q_i,q_j)=1$ if $q_i$ and $q_j$ interact at timestep $m$ and zero otherwise. Furthermore, if $q_i$ and $q_j$ interact at timestep $t$, then $w_t(q_i,q_j) = \infty$. In the second phase of the FGP-OEE method, the Overall Extreme Exchange \cite{park1995algorithms} graph partitioning algorithm is applied to the weighted graph of each timestep $t$ starting from the partitioning of the previous timestep $t-1$. A random valid partitioning (\ie interacting qubits are in the same core) is selected for the first timeslice. In the relaxed version of the method the OEE iterative partitioning optimization algorithm for each timeslice is stopped as soon as the qubit partitioning is valid.

For a random circuit with 200 slices, the proposed technique achieve a $28.26\%$ reduction in terms of inter-core communications with respect to the baseline. For the QNN and Cuccaro Adders workloads we measured a $48.82\%$ and $40.53\%$ reduction respectively. Modest improvements were measured for the Quantum Volume and Graph State circuits. We also report a degradation ranging from $-47.08\%$ to $-32.44\%$ for highly structured circuits such as the Draper Adder and the QFT. This result, is significant in that training the model on randomly generated circuits only is not enough to achieve optimized solutions for highly structured circuits. In this regard, recent efforts in synthetic circuits dataset generation could be impactful \cite{apak2024ketgpt}.

\section{Related Works}

Several works have addressed the qubit allocation problem for single-core architectures with optimal and heuristic approaches  \cite{lin2023scalable, li2019tackling, nannicini2022optimal}. In addressing the multi-core qubit allocation challenge Baker et al. \cite{baker2020time} introduced the FGP-OEE heuristic algorithm for paritioning qubits of a time-sliced circuit. A QUBO formulation was introduced in the work by Bandic et al. \cite{bandic2023mapping}. While envisioning a future resolution of this problem through quantum annealers, the current complexity lies in the tuning of the parameter tuning balancing feasibility and optimality in the objective, which can be a challenging and time-consuming task. Escofet et al. \cite{escofet2023hungarian,escofet2024revisiting} proposed a solution based on the Hungarian algorithm for assignments problems and derived theoretical bounds for the amount of inter-core state-transfers in all-to-all connected modular architectures. Cuomo et al. \cite{cuomo2023optimized} introduced a formal problem formulation of the quantum compiling problem in distributed quantum systems focusing on remote gates, proposing a dynamic network flow model to minimize inter-core gates. Zhang et al. \cite{zhang2023compilation} introduced compilation for quantum chiplets considering inter-core communication through an \textit{highway} of entangled qubits.

%\cite{yu2022gnn, pastor2023towards}

\section{Conclusion}

In our discussion, we formulated the problem assuming communication through state transfers, as often occurs in multi-core architectures. The same approach described in this work, namely a policy that provides qubit assignments for each slice, could be adapted to single-core cases or, in any case, multi-core scenarios where state movement occurs through swaps. In such cases, the problem becomes more complex, particularly in the calculation of the reward. In the single-core case, for example, it would be necessary to calculate, for each slice, the minimum number of swaps required to transition from one permutation to another, taking into account the system's topology. This problem is known as Token Swapping and remains $W[1]$-hard \cite{bonnet2018complexity}.

%A more advanced masking mechanism could could be investigated to deal with cores featuring physical qubits with sparse connectivity. In this work, we considered the core capacity to be the number of free qubits, while in such a scenario it could be considered to be the amount of free intra-core topology edges. At each decoding step, all the edges having the action qubit as one of the vertexes would be masked for subsequent decoding steps and the edge capacity for the core will decrease.

In future works, a more advanced masking mechanism could could be investigated to deal with cores featuring sparsely connected physical qubits, considering the amount of available edges as capacities instead of free qubits. More advanced policy training algorithms, such as Proximal Policy Optimization (PPO), could potentially lead to better performance by avoiding local minima. Additionally, the policy could be enhanced by introducing techniques such as relative positional encoding of slices \cite{shaw2018self}, taking into consideration the intrinsic symmetry of quantum circuits. Regarding the training dataset, circuits resembling real highly structured algorithms could be included rather than considering only randomly generated circuits. 

Training attention-based models is computational intensive but prone to parallel execution \cite{vaswani2017attention}. With the availability of increased computational resources, the scalability of the proposed approach to a higher amount of qubits remains to be assessed. Given the promising results about the transformer scalability in context of several thousands of words in the Natural Language Domain, there are reasons to be optimistic about the capability for the proposed attention-based model to scale for thousands of circuit slices.

In conclusion, this work investigated the challenges associated with achieving scalability in quantum computing systems. The imperative need to optimize communication between cores and minimize state transfers while adhering to architectural constraints underscores the complexity of the compilation and mapping of quantum circuits onto physical qubits. Addressing the NP-hard nature of the compilation process, this work has proposed a novel approach employing DRL methods to learn heuristic solutions tailored to a specific multi-core architecture. The experimental evaluations have demonstrated the effectiveness of the proposed method in outperforming baseline approaches, showcasing its ability to reduce inter-core communications and minimize online time-to-solution which we believe makes DRL a promising near-term approach for qubit allocation in modular quantum architectures.

%%---------------------------------------------------------------------------------
\section*{Acknowledgement}
Authors gratefully acknowledge funding from the European Commission through HORIZON-EIC-2022-PATHFINDEROPEN01-101099697 (QUADRATURE)

%\section*{Acknowledgements}
%This document is derived from previous conferences, in particular ISCA 2024 and HPCA 2021.

%%%%%%% -- PAPER CONTENT ENDS -- %%%%%%%%

%%%%%%%%% -- BIB STYLE AND FILE -- %%%%%%%%
\bibliographystyle{IEEEtranS}
\bibliography{refs}

% Generated by IEEEtranS.bst, version: 1.13 (2008/09/30)
\begin{thebibliography}{10}
\providecommand{\url}[1]{#1}
\csname url@samestyle\endcsname
\providecommand{\newblock}{\relax}
\providecommand{\bibinfo}[2]{#2}
\providecommand{\BIBentrySTDinterwordspacing}{\spaceskip=0pt\relax}
\providecommand{\BIBentryALTinterwordstretchfactor}{4}
\providecommand{\BIBentryALTinterwordspacing}{\spaceskip=\fontdimen2\font plus
\BIBentryALTinterwordstretchfactor\fontdimen3\font minus \fontdimen4\font\relax}
\providecommand{\BIBforeignlanguage}[2]{{%
\expandafter\ifx\csname l@#1\endcsname\relax
\typeout{** WARNING: IEEEtranS.bst: No hyphenation pattern has been}%
\typeout{** loaded for the language `#1'. Using the pattern for}%
\typeout{** the default language instead.}%
\else
\language=\csname l@#1\endcsname
\fi
#2}}
\providecommand{\BIBdecl}{\relax}
\BIBdecl

\bibitem{aaronson2013quantum}
S.~Aaronson, \emph{Quantum computing since Democritus}.\hskip 1em plus 0.5em minus 0.4em\relax Cambridge University Press, 2013.

\bibitem{apak2024ketgpt}
B.~Apak, M.~Bandic, A.~Sarkar, and S.~Feld, ``Ketgpt--dataset augmentation of quantum circuits using transformers,'' \emph{arXiv preprint arXiv:2402.13352}, 2024.

\bibitem{arute2019quantum}
F.~Arute, K.~Arya, R.~Babbush, D.~Bacon, J.~C. Bardin, R.~Barends, R.~Biswas, S.~Boixo, F.~G. Brandao, D.~A. Buell \emph{et~al.}, ``Quantum supremacy using a programmable superconducting processor,'' \emph{Nature}, vol. 574, no. 7779, pp. 505--510, 2019.

\bibitem{baker2020time}
J.~M. Baker, C.~Duckering, A.~Hoover, and F.~T. Chong, ``Time-sliced quantum circuit partitioning for modular architectures,'' in \emph{Proceedings of the 17th ACM International Conference on Computing Frontiers}, 2020, pp. 98--107.

\bibitem{bandic2023mapping}
M.~Bandic, L.~Prielinger, J.~N{\"u}{\ss}lein, A.~Ovide, S.~Rodrigo, S.~Abadal, H.~van Someren, G.~Vardoyan, E.~Alarcon, C.~G. Almudever \emph{et~al.}, ``Mapping quantum circuits to modular architectures with qubo,'' in \emph{2023 IEEE International Conference on Quantum Computing and Engineering (QCE)}, vol.~1.\hskip 1em plus 0.5em minus 0.4em\relax IEEE, 2023, pp. 790--801.

\bibitem{barenco1995elementary}
A.~Barenco, C.~H. Bennett, R.~Cleve, D.~P. DiVincenzo, N.~Margolus, P.~Shor, T.~Sleator, J.~A. Smolin, and H.~Weinfurter, ``Elementary gates for quantum computation,'' \emph{Physical review A}, vol.~52, no.~5, p. 3457, 1995.

\bibitem{bello2016neural}
I.~Bello, H.~Pham, Q.~V. Le, M.~Norouzi, and S.~Bengio, ``Neural combinatorial optimization with reinforcement learning,'' \emph{arXiv preprint arXiv:1611.09940}, 2016.

\bibitem{berto2023rl4co}
F.~Berto, C.~Hua, J.~Park, M.~Kim, H.~Kim, J.~Son, H.~Kim, J.~Kim, and J.~Park, ``{RL}4{CO}: a unified reinforcement learning for combinatorial optimization library,'' in \emph{NeurIPS 2023 Workshop: New Frontiers in Graph Learning}, 2023, \url{https://github.com/ai4co/rl4co}.

\bibitem{bonnet2018complexity}
{\'E}.~Bonnet, T.~Miltzow, and P.~Rzazewski, ``Complexity of token swapping and its variants,'' \emph{Algorithmica}, vol.~80, pp. 2656--2682, 2018.

\bibitem{botea2018complexity}
A.~Botea, A.~Kishimoto, and R.~Marinescu, ``On the complexity of quantum circuit compilation,'' in \emph{Proceedings of the International Symposium on Combinatorial Search}, vol.~9, no.~1, 2018, pp. 138--142.

\bibitem{chang2023attention}
J.~Chang, J.~Wang, B.~Li, Y.~Zhao, and D.~Li, ``Attention-based deep reinforcement learning for edge user allocation,'' \emph{IEEE Transactions on Network and Service Management}, 2023.

\bibitem{charbon2020cryogenic}
E.~Charbon, M.~Babaie, A.~Vladimirescu, and F.~Sebastiano, ``Cryogenic cmos circuits and systems: Challenges and opportunities in designing the electronic interface for quantum processors,'' \emph{IEEE Microwave Magazine}, vol.~22, no.~1, pp. 60--78, 2020.

\bibitem{chen2021decision}
L.~Chen, K.~Lu, A.~Rajeswaran, K.~Lee, A.~Grover, M.~Laskin, P.~Abbeel, A.~Srinivas, and I.~Mordatch, ``Decision transformer: Reinforcement learning via sequence modeling,'' \emph{Advances in neural information processing systems}, vol.~34, pp. 15\,084--15\,097, 2021.

\bibitem{chen2022deep}
R.~Chen, W.~Li, and H.~Yang, ``A deep reinforcement learning framework based on an attention mechanism and disjunctive graph embedding for the job-shop scheduling problem,'' \emph{IEEE Transactions on Industrial Informatics}, vol.~19, no.~2, pp. 1322--1331, 2022.

\bibitem{cuomo2023optimized}
D.~Cuomo, M.~Caleffi, K.~Krsulich, F.~Tramonto, G.~Agliardi, E.~Prati, and A.~S. Cacciapuoti, ``Optimized compiler for distributed quantum computing,'' \emph{ACM Transactions on Quantum Computing}, vol.~4, no.~2, pp. 1--29, 2023.

\bibitem{droste2002analysis}
S.~Droste, T.~Jansen, and I.~Wegener, ``On the analysis of the (1+ 1) evolutionary algorithm,'' \emph{Theoretical Computer Science}, vol. 276, no. 1-2, pp. 51--81, 2002.

\bibitem{escofet2023hungarian}
P.~Escofet, A.~Ovide, C.~G. Almudever, E.~Alarc{\'o}n, and S.~Abadal, ``Hungarian qubit assignment for optimized mapping of quantum circuits on multi-core architectures,'' \emph{IEEE Computer Architecture Letters}, 2023.

\bibitem{escofet2024revisiting}
P.~Escofet, A.~Ovide, M.~Bandic, L.~Prielinger, H.~van Someren, S.~Feld, E.~Alarc{\'o}n, S.~Abadal, and C.~G. Almud{\'e}ver, ``Revisiting the mapping of quantum circuits: Entering the multi-core era,'' \emph{ACM Transactions on Quantum Computing}, 2024.

\bibitem{pytorch_geometric}
M.~Fey and J.~E. Lenssen, ``Fast graph representation learning with {PyTorch Geometric},'' in \emph{ICLR Workshop on Representation Learning on Graphs and Manifolds}, 2019.

\bibitem{gen1997genetic}
M.~Gen, R.~Cheng, and D.~Wang, ``Genetic algorithms for solving shortest path problems,'' in \emph{Proceedings of 1997 IEEE International Conference on Evolutionary Computation (ICEC'97)}.\hskip 1em plus 0.5em minus 0.4em\relax IEEE, 1997, pp. 401--406.

\bibitem{giannelli2022tutorial}
L.~Giannelli, P.~Sgroi, J.~Brown, G.~S. Paraoanu, M.~Paternostro, E.~Paladino, and G.~Falci, ``A tutorial on optimal control and reinforcement learning methods for quantum technologies,'' \emph{Physics Letters A}, vol. 434, p. 128054, 2022.

\bibitem{grover1996fast}
L.~K. Grover, ``A fast quantum mechanical algorithm for database search,'' in \emph{Proceedings of the twenty-eighth annual ACM symposium on Theory of computing}, 1996, pp. 212--219.

\bibitem{gurobi}
\BIBentryALTinterwordspacing
{Gurobi Optimization, LLC}, ``{Gurobi Optimizer Reference Manual},'' 2023. [Online]. Available: \url{https://www.gurobi.com}
\BIBentrySTDinterwordspacing

\bibitem{harik1999compact}
G.~R. Harik, F.~G. Lobo, and D.~E. Goldberg, ``The compact genetic algorithm,'' \emph{IEEE transactions on evolutionary computation}, vol.~3, no.~4, pp. 287--297, 1999.

\bibitem{jnane2022multicore}
H.~Jnane, B.~Undseth, Z.~Cai, S.~C. Benjamin, and B.~Koczor, ``Multicore quantum computing,'' \emph{Physical Review Applied}, vol.~18, no.~4, p. 044064, 2022.

\bibitem{kim2023devformer}
H.~Kim, M.~Kim, F.~Berto, J.~Kim, and J.~Park, ``Devformer: A symmetric transformer for context-aware device placement,'' in \emph{International Conference on Machine Learning}.\hskip 1em plus 0.5em minus 0.4em\relax PMLR, 2023, pp. 16\,541--16\,566.

\bibitem{kipf2016semi}
T.~N. Kipf and M.~Welling, ``Semi-supervised classification with graph convolutional networks,'' \emph{arXiv preprint arXiv:1609.02907}, 2016.

\bibitem{kool2018attention}
W.~Kool, H.~Van~Hoof, and M.~Welling, ``Attention, learn to solve routing problems!'' \emph{arXiv preprint arXiv:1803.08475}, 2018.

\bibitem{li2019tackling}
G.~Li, Y.~Ding, and Y.~Xie, ``Tackling the qubit mapping problem for nisq-era quantum devices,'' in \emph{Proceedings of the Twenty-Fourth International Conference on Architectural Support for Programming Languages and Operating Systems}, 2019, pp. 1001--1014.

\bibitem{li2023survey}
W.~Li, H.~Luo, Z.~Lin, C.~Zhang, Z.~Lu, and D.~Ye, ``A survey on transformers in reinforcement learning,'' \emph{arXiv preprint arXiv:2301.03044}, 2023.

\bibitem{lin2023scalable}
W.-H. Lin, J.~Kimko, B.~Tan, N.~Bj{\o}rner, and J.~Cong, ``Scalable optimal layout synthesis for nisq quantum processors,'' in \emph{2023 60th ACM/IEEE Design Automation Conference (DAC)}.\hskip 1em plus 0.5em minus 0.4em\relax IEEE, 2023, pp. 1--6.

\bibitem{montanaro2016quantum}
A.~Montanaro, ``Quantum algorithms: an overview,'' \emph{npj Quantum Information}, vol.~2, no.~1, pp. 1--8, 2016.

\bibitem{nannicini2022optimal}
G.~Nannicini, L.~S. Bishop, O.~G{\"u}nl{\"u}k, and P.~Jurcevic, ``Optimal qubit assignment and routing via integer programming,'' \emph{ACM Transactions on Quantum Computing}, vol.~4, no.~1, pp. 1--31, 2022.

\bibitem{nielsen2010quantum}
M.~A. Nielsen and I.~L. Chuang, \emph{Quantum computation and quantum information}.\hskip 1em plus 0.5em minus 0.4em\relax Cambridge university press, 2010.

\bibitem{nowling2011priority}
R.~J. Nowling and H.~Mauch, ``Priority encoding scheme for solving permutation and constraint problems with genetic algorithms and simulated annealing,'' in \emph{2011 Eighth International Conference on Information Technology: New Generations}.\hskip 1em plus 0.5em minus 0.4em\relax IEEE, 2011, pp. 810--815.

\bibitem{ovide2023mapping}
A.~Ovide, S.~Rodrigo, M.~Bandic, H.~Van~Someren, S.~Feld, S.~Abadal, E.~Alarcon, and C.~G. Almudever, ``Mapping quantum algorithms to multi-core quantum computing architectures,'' in \emph{2023 IEEE International Symposium on Circuits and Systems (ISCAS)}, 2023, pp. 1--5.

\bibitem{park1995algorithms}
T.~Park and C.~Y. Lee, ``Algorithms for partitioning a graph,'' \emph{Computers \& Industrial Engineering}, vol.~28, no.~4, pp. 899--909, 1995.

\bibitem{pytorch}
A.~Paszke, S.~Gross, F.~Massa, A.~Lerer, J.~Bradbury, G.~Chanan, T.~Killeen, Z.~Lin, N.~Gimelshein, L.~Antiga \emph{et~al.}, ``Pytorch: An imperative style, high-performance deep learning library,'' \emph{Advances in neural information processing systems}, vol.~32, 2019.

\bibitem{Qiskit}
{Qiskit contributors}, ``Qiskit: An open-source framework for quantum computing,'' 2023.

\bibitem{quetschlich2023mqtbench}
N.~Quetschlich, L.~Burgholzer, and R.~Wille, ``Mqt bench: Benchmarking software and design automation tools for quantum computing,'' \emph{Quantum}, vol.~7, p. 1062, 2023.

\bibitem{rodrigo2021modelling}
S.~Rodrigo, S.~Abadal, C.~G. Almud{\'e}ver, and E.~Alarc{\'o}n, ``Modelling short-range quantum teleportation for scalable multi-core quantum computing architectures,'' in \emph{Proceedings of the Eight Annual ACM International Conference on Nanoscale Computing and Communication}, 2021, pp. 1--7.

\bibitem{sarovar2020detecting}
M.~Sarovar, T.~Proctor, K.~Rudinger, K.~Young, E.~Nielsen, and R.~Blume-Kohout, ``Detecting crosstalk errors in quantum information processors,'' \emph{Quantum}, vol.~4, p. 321, 2020.

\bibitem{shaw2018self}
P.~Shaw, J.~Uszkoreit, and A.~Vaswani, ``Self-attention with relative position representations,'' \emph{arXiv preprint arXiv:1803.02155}, 2018.

\bibitem{shor1999polynomial}
P.~W. Shor, ``Polynomial-time algorithms for prime factorization and discrete logarithms on a quantum computer,'' \emph{SIAM review}, vol.~41, no.~2, pp. 303--332, 1999.

\bibitem{siraichi2018qubit}
M.~Y. Siraichi, V.~F.~d. Santos, C.~Collange, and F.~M.~Q. Pereira, ``Qubit allocation,'' in \emph{Proceedings of the 2018 International Symposium on Code Generation and Optimization}, 2018, pp. 113--125.

\bibitem{smith2022scaling}
K.~N. Smith, G.~S. Ravi, J.~M. Baker, and F.~T. Chong, ``Scaling superconducting quantum computers with chiplet architectures,'' in \emph{2022 55th IEEE/ACM International Symposium on Microarchitecture (MICRO)}.\hskip 1em plus 0.5em minus 0.4em\relax IEEE, 2022, pp. 1092--1109.

\bibitem{sutton2018reinforcement}
R.~S. Sutton and A.~G. Barto, \emph{Reinforcement learning: An introduction}.\hskip 1em plus 0.5em minus 0.4em\relax MIT press, 2018.

\bibitem{vaswani2017attention}
A.~Vaswani, N.~Shazeer, N.~Parmar, J.~Uszkoreit, L.~Jones, A.~N. Gomez, {\L}.~Kaiser, and I.~Polosukhin, ``Attention is all you need,'' \emph{Advances in neural information processing systems}, vol.~30, 2017.

\bibitem{vinyals2015order}
O.~Vinyals, S.~Bengio, and M.~Kudlur, ``Order matters: Sequence to sequence for sets,'' \emph{arXiv preprint arXiv:1511.06391}, 2015.

\bibitem{vinyals2015pointer}
O.~Vinyals, M.~Fortunato, and N.~Jaitly, ``Pointer networks,'' \emph{Advances in neural information processing systems}, vol.~28, 2015.

\bibitem{wang2020deep}
H.-n. Wang, N.~Liu, Y.-y. Zhang, D.-w. Feng, F.~Huang, D.-s. Li, and Y.-m. Zhang, ``Deep reinforcement learning: a survey,'' \emph{Frontiers of Information Technology \& Electronic Engineering}, vol.~21, no.~12, pp. 1726--1744, 2020.

\bibitem{williams1992simple}
R.~J. Williams, ``Simple statistical gradient-following algorithms for connectionist reinforcement learning,'' \emph{Machine learning}, vol.~8, pp. 229--256, 1992.

\bibitem{zhang2023compilation}
H.~Zhang, K.~Yin, A.~Wu, H.~Shapourian, A.~Shabani, and Y.~Ding, ``Compilation for quantum computing on chiplets,'' \emph{arXiv preprint arXiv:2305.05149}, 2023.

\end{thebibliography}
%%%%%%%%%%%%%%%%%%%%%%%%%%%%%%%%%%%%

\end{document}